\documentclass[nohyper,letterpaper]{JHEP3}

\usepackage{epsfig,multicol,bbm}
\usepackage{mathrsfs,amsmath,amssymb,graphicx,eucal}

\newcommand\fverb{\setbox\fverbbox=\hbox\bgroup\verb}
\newcommand\fverbdo{\egroup\medskip\noindent\fbox{\unhbox\fverbbox}\ }
\newcommand\fverbit{\egroup\item[\fbox{\unhbox\fverbbox}]}
\newbox\fverbbox

\newcommand{\de}{{\mathrm d}}
\newcommand{\om}{{\Omega_m}}
\newcommand{\ob}{{\Omega_b}}
\newcommand{\odm}{{\Omega_\mathrm{DM}}}

\newcommand{\ol}{{\Omega_\Lambda}}
\newcommand{\ho}{{H_0}}
\newcommand{\lcdm}{$\Lambda$CDM}
\newcommand{\nhs}{{n_\mathrm{HS}}}
\newcommand{\nst}{{n_\mathrm{St}}}

\title{Tomography from the Next Generation of Cosmic Shear Experiments for Viable $f(R)$ Models}

\author{Stefano Camera,$^{a,b,c}$ Antonaldo Diaferio$^{a,c,d}$ and Vincenzo F. Cardone$^e$\\$^a$Dipartimento di Fisica Generale ``A. Avogadro'', Universit\`a degli Studi di Torino, via P. Giuria 1, 10125 Torino, Italy\\$^b$Dipartimento di Fisica Teorica, Universit\`a degli Studi di Torino, via P. Giuria 1, 10125 Torino, Italy\\$^c$Istituto Nazionale di Fisica Nucleare (INFN), Sezione di Torino, via P. Giuria 1, 10125 Torino, Italy\\$^d$Harvard-Smithsonian Center for Astrophysics, 60 Garden St., Cambridge, MA 02138, USA\\$^e$Istituto Nazionale di Astrofisica (INAF), Osservatorio Astronomico di Roma, via Frascati 33, 00040 Monteporzio Catone, Italy\\E-mail: \email{camera@ph.unito.it, diaferio@ph.unito.it, winnyenodrac@gmail.com}}

\preprint{\today}

\abstract{We present the cosmic shear signal predicted by two viable cosmological models in the framework of modified-action $f(R)$ theories. We use $f(R)$ models where the current accelerated expansion of the Universe is a direct consequence of the modified gravitational Lagrangian rather than Dark Energy (DE), either in the form of vacuum energy/cosmological constant or of a dynamical scalar field (e.g. quintessence). We choose Starobinsky's (St) and Hu \& Sawicki's (HS) $f(R)$ models, which are carefully designed to pass the Solar System gravity tests. In order to further support -- or rule out -- $f(R)$ theories as alternative candidates to the DE hypothesis, we exploit the power of weak gravitational lensing, specifically of cosmic shear. We calculate the tomographic shear matrix as it would be measured by the upcoming ESA Cosmic Vision Euclid satellite. We find that in the St model the cosmic shear signal is almost completely degenerate with \lcdm, but it is easily distinguishable in the HS model. Moreover, we compute the corresponding Fisher matrix for both the St and HS models, thus obtaining forecasts for their cosmological parameters. Finally, we show that the Bayes factor for cosmic shear will definitely favour the HS model over \lcdm\ if Euclid measures a value larger than $\sim0.02$ for the extra HS parameter $\nhs$.}


\begin{document}

\section{Introduction}
It is now widely accepted that a number of different cosmological observations points out a lack in our current understanding of the Universe. In particular, the temperature anisotropy spectrum of the Cosmic Microwave Background (CMB) \cite{deBernardis:2000gy,Stompor:2001xf,Netterfield:2001yq,Rebolo:2004vp,Bennett:2003bz,Spergel:2003cb,Komatsu:2008hk}, the Hubble diagram of Type Ia Supernov\ae~(SNeIa) \cite{Perlmutter:1996ds,Riess:1998cb,Schmidt:1998ys,Perlmutter:1998np,Knop:2003iy,Tonry:2003zg,Riess:2004n,Astier:2005qq,WoodVasey:2007jb} and the clustering properties probing the cosmic Large-Scale Structure (LSS) \cite{Percival:2002gq,Pope:2004cc} are concordant pieces of evidence in favour of a present period of accelerated expansion of the Universe. In the current cosmological model $\Lambda$ Cold Dark Matter ($\Lambda$CDM), this observational evidence and the problem of the missing mass in the dynamics of galaxies and galaxy clusters and in the LSS of the Universe \cite{Zwicky:1933gu,Zwicky:1937zza,Dodelson:2001ux,Hawkins:2002sg,Spergel:2006hy,Riess:2006fw} are excellently reproduced and solved by assuming a spatially flat Universe dominated by cold Dark Matter (DM) and a vacuum Dark Energy (DE) in form of a cosmological constant $\Lambda$.

However, this scenario has serious drawbacks: if $\Lambda$ is interpreted as vacuum energy, its value is $120$ orders of magnitude smaller than what is expected from quantum field theory; the coincidence and fine-tuning problems do not seem to have any natural explanation and DE has to represent $\sim70\%$ of the total energy budget of the Universe. This circumstance has motivated the search for alternative solutions. They mostly rely on scalar fields with suitable potentials which provide a varying $\Lambda$ term or other DE components or even unified DM and DE fluids \cite{Peebles:2002gy,Padmanabhan:2002ji,Copeland:2006wr,2010deto.book.....A,2010AdAst2010E..78B,Sapone:2010iz,Li:2011sd} with exotic properties.

Recently, a different route has started to be followed. The cosmological constant was originally proposed as a geometric term, i.e. a constant term in the left-hand (geometric) side of the Einstein equation, thus adding no extra component in the stress-energy tensor of the fluid filling the Universe. By generalising this approach, one can argue whether it is possible to reproduce the current cosmic accelerated expansion by adding a non-constant time-dependent term in Einstein's tensor. The effort of modifying and generalising the Hilbert-Einstein action of gravity actually dates back to just few years after Einstein's seminal papers (e.g. \cite{Schmidt:2006jt} for a historical review), and it has been also proposed by Starobinsky \cite{Starobinsky:1980te} in order to explain the cosmic inflation in the early Universe. This idea has been suggested again nowadays for the purpose of correctly describing the current accelerated expansion of the Universe without any exotic fluid \cite{Capozziello:2002rd,Capozziello:2003gx,Nojiri:2003ft,Nojiri:2003wx,Carroll:2003wy,Capozziello:2005ku}. These modified-action theories of gravity are widely known as $f(R)$ theories, because in the gravity Lagrangian the Ricci scalar $R$ is replaced by a generic function $f(R)$.

However, GR is a well-tested theoretical framework, at least with respect to the Solar System scale of distances. Therefore, any $f(R)$ theory which attempts to solve the late-time acceleration problem has to face the Solar System tests of gravity. Recently, two models carefully designed to pass the local gravity tests but still providing an accelerated cosmic expansion have been proposed \cite{Starobinsky:2007hu,Hu:2007nk}.

In this paper we use weak gravitational lensing to test and constrain these models. Indeed, weak lensing is a powerful tool \cite{Kaiser:1991qi,Heavens:1994iq,Kaiser:1996tp,Hu:1998az,Hu:1999ek,Hu:2000ee,Hu:2001fb,Schmidt:2008hc,Thomas:2008tp}: gravitational lens effects are due to the deflection of light occuring when photons travel near matter, i.e. in the presence of a non-negligible gravitational field. The cosmic convergence and shear encapsulate information about both the source emitting light and the structures that photons cross before arriving at the telescope. Therefore, weak lensing allows to explore both the basis of the cosmological model and LSS of the Universe, in other words it brings information about the geometry and the dynamics. Hence, the study of the power spectrum of weak lensing can be a crucial test.

The structure of this work is as follows. In \S~\ref{sec:theory}, we present the theory of $f(R)$ models, both at background level and in the linear theory of cosmological perturbations. In \S~\ref{sec:wl}, we recast the most important observables of weak gravitational lensing, the convergence and the cosmic shear, and their power spectra. Then, we also outline how to construct the tomographic shear matrix. \S~\ref{sec:fisher} illustrates the cosmic shear Fisher matrix, whilst \S~\ref{ssec:B-evidence} the Bayes factor and its use in the framework of model selection. Finally, we show our results in \S~\ref{sec:results}. In \S~\ref{sec:conclusions}, conclusions are drawn.

\section{$f(R)$ Cosmology}\label{sec:theory}
In modified-action theories of gravity, the standard Hilbert-Einstein Lagrangian density of GR in the $\Lambda$CDM model
\begin{equation}
\mathscr L_\mathrm{HE}=\frac{R-2\Lambda}{16\pi G}
\end{equation}
is replaced by a more general function $f(R)$ of the Ricci scalar $R$ (see \cite{Sotiriou:2008rp} for an exhaustive review),
\begin{equation}
\mathscr L=\frac{f(R)}{16\pi G}.
\end{equation}

Modified Einstein's equations are given by the variation of the Lagrangian with respect to the metric. This gives, after some manipulations and modulo surface terms,\footnote{We use units such that $c=1$ and signature $\{-,+,+,+\}$, where Greek indeces run over spacetime dimension, whereas Latin indeces label spatial coordinates.}
\begin{equation}
f_{,R}R_{\mu\nu}-\nabla_{\mu\nu}f_{,R}+\left(\Box f_{,R}-\frac{1}{2}f\right)g_{\mu\nu}=8\pi GT_{\mu\nu}\label{modEin},
\end{equation}
where $_{,R}$ denotes a derivative with respect to $R$, $R_{\mu\nu}$ is the Ricci tensor, $g_{\mu\nu}$ the spacetime metric and $T_{\mu\nu}$ the energy-momentum tensor.

In these modified theories of gravity, the usual massless spin-2 graviton is not the only carrier of the gravitational interaction. Indeed, there is also a scalar degree of freedom -- often dubbed scalaron --, which can be conveniently described by the function $\phi=f_{,R}-1$. The evolution of the field $\phi$ obeys to the trace of Eq.~\eqref{modEin}
\begin{equation}
\Box\phi=\frac{\de V}{\de\phi}+\frac{8\pi G}{3}T,
\end{equation}
where the potential $V$ is related to $R$ by
\begin{equation}
\frac{\de V}{\de R}=\frac{1}{3}\left(2f-Rf_{,R}\right)f_{,RR},
\end{equation}
and $T$ is the trace of the stress-energy tensor.

Albeit the choice of the functional form of $f(R)$ is in principle completely free, there are some constraints which help to avoid such an arbitrariness. First, modifying the gravity Lagrangian leads to deviations from GR at all scales. Particularly, the presence of the scalar degree of freedom $\phi$, which couples to matter, is responsible for a long-range fifth force. At the scale of the Solar System, this new force can possibly lead to wrong values of the PPN parameters \cite{Dominguez:2004ds,Olmo:2005jd,Sotiriou:2005xe}. A usual way to solve similar problems is to reassociate high densities with high curvatures, so that $\phi$ becomes very massive and the fifth force escapes any detection. Such a behaviour of the model, namely a large and positive mass squared term $m^2=(8\pi G)^2\rho/3$ in high curvature environments, is usually called ``chameleon effect.'' On the other hand, the Big-Bang nucleosynthesis and early Universe data, for example the temperature anisotropies of the CMB, strictly constrain the choice of the cosmological model. In other words, one wants to recover GR at high redshift $z$. Similarly, the background evolution of the cosmos should not be too different from what predicted by the \lcdm\ model, since this model reproduces SNeIa data excellently.

The constraints which must be satisfied by $f(R)$ can be summarised as
\begin{align}
\lim_{R\to0}f(R)&=0,\\
\lim_{R\to\infty}f(R)&=\mathrm{const.},\\
\left.f_{,R}\right|_{R\gg m^2}&>0.
\end{align}
A further condition which ensures that the solution is stable at high curvatures should be included. This translates into
\begin{equation}
\left.f_{,RR}\right|_{R\gg m^2}>0.
\end{equation}

Among the possible choices left out by the above conditions, in this paper we will consider two popular classes of $f(R)$ models which we briefly describe in the following, namely the models proposed by \cite{Starobinsky:2007hu} and \cite{Hu:2007nk}.

\paragraph{Starobinsky}
Starobinsky's model \cite{Starobinsky:2007hu} (St, hereafter) is described by the function
\begin{equation}
f(R)=R+\lambda R_\star\left[\left(1+\frac{R^2}{{R_\star}^2}\right)^{-\nst}-1\right],
\end{equation}
with $R_\star$ a scaling curvature parameter and $\lambda$ and $\nst$ two positive dimensionless constants. It is worth noting that there is no cosmological constant, because $f(0)=0$. Nonetheless, in high curvature r\'egime, one recovers an effective $\Lambda$-like term, since $f(R\gg R_\star)\simeq R-2\Lambda^{(\mathrm{eff})}$, where $\Lambda^{(\mathrm{eff})}=\lambda R_\star/2$. Moreover, in the early Universe the Hubble parameter $H$ is the same as in a \lcdm~model with an effective matter fraction $\Omega_m^{(\mathrm{eff})}=1-\lambda R_\star/(6\ho^2)$, where $\ho=100h\,\mathrm{km~s^{-1}~Mpc^{-1}}$ is the Hubble constant. This guarantees that the Big-Bang nucleosynthesis constraints are satisfied.

\paragraph{Hu \& Sawicki}
Hu \& Sawicki \cite{Hu:2007nk} proposed an $f(R)$ model (HS, hereafter) described by
\begin{equation}
f(R)=R-m^2\frac{c_1\left(R/m^2\right)^\nhs}{1+c_2\left(R/m^2\right)^\nhs},
\end{equation}
where $c_1$, $c_2$ and $\nhs$ are positive dimensionless constants. As in the St model, $f(0)=0$, thus there is no formal cosmological constant. However, in high-curvature environments, namely when $m^2/R\to0$, an effective $\Lambda$-like term is present, since
\begin{equation}
\lim_{m^2/R\to0}f(R)\simeq-\frac{c_1}{c_2}m^2+\frac{c_1}{{c_2}^2}m^2\left(\frac{m^2}{R}\right)^\nhs.
\end{equation}
Finally, in the HS model the expansion history $H$ in the early Universe is again the same as in the \lcdm~model. In the present case, the effective matter fraction is given by $\Omega_m^{(\mathrm{eff})}=6c_2/(c_1+6c_2)$.
%\cite{Cardone:2010,Camera:2009uz}

\subsection{Background Evolution}
In a spatially flat Friedmann-Lema\^itre-Robertson-Wal\-ker (FLRW) Universe with length element
\begin{equation}
\overline g_{\mu\nu}\,\de x^\mu\de x^\nu=-\de t^2+a^2(t)\de\mathbf x^2\label{flatFLRW}
\end{equation}
filled with a perfect fluid with energy density $\rho$ and pressure $p$, the modified Einstein's equations \eqref{modEin} give a modified Friedmann's equation for the scale factor $a=1/(1+z)$. By writing Eq.~\eqref{modEin} in terms of the Hubble parameter $H=\de\ln a/\de t$, the Friedmann equation becomes
\begin{equation}
H^2+\frac{\dot{f_{,R}}}{f_{,R}}H+\frac{f-Rf_{,R}}{6f_{,R}}=\frac{8\pi G}{3f_{,R}}\rho\label{modFri},
\end{equation}
where a dot denotes a derivative with respect to the cosmic time $t$. We remind the reader that in a FLRW Universe, the Ricci tensor can be written as
\begin{equation}
R=6\left(\dot H+2H^2\right).
\end{equation}

Cardone et al. \cite{Cardone:2010} solved the modified Friedmann equation \eqref{modFri} for a large set of randomly selected parameters $\{\om$, $\lambda/c_1$, $R_\star/c_2$, $\nst/\nhs\}$, and fit the expansion history of the Universe with a number of updated and accurate data. Specifically, they fitted the background expansion of the cosmos using the Hubble diagram of SneIa \cite{Hicken:2009dk} and Gamma Ray Bursts \cite{Cardone:2010}. Moreover, they use $H(z)$ data from passively evolving red galaxies \cite{Freedman:2000cf}, Baryon Acoustic Oscillations extracted from the seventh data release of the Sloan Digital Sky Survey, and distance priors from the recent WMAP7 data \cite{Larson:2010gs}. They found that the Hubble parameter can be excellently approximated by
\begin{equation}
\frac{H(z)}{\ho}=\left\{\begin{array}{lr}
\mathcal E(z)E_\mathrm{CPL}(z)+\left[1-\mathcal E(z)\right]E_\Lambda(z)&z\leq z_\Lambda\\
\sqrt{\Omega_m^{(\mathrm{eff})}\left(1+z\right)^3+1-\Omega_m^{(\mathrm{eff})}}&z>z_\Lambda
\end{array}\right.,\label{H_z}
\end{equation}
where
\begin{equation}
\mathcal E(z)=\sum_{i=1}^3e_i\left(z-z_\Lambda\right)^i
\end{equation}
is an interpolating function, with $e_i$ and $z_\Lambda$ fitting parameters, and
\begin{equation}
E_\mathrm{CPL}=\om\left(1+z\right)^3+\left(1-\om\right)\left(1+z\right)^{3(1+w_0+w_a)}e^{-\frac{3w_az}{1+z}}\label{cpl}
\end{equation}
is the dimensionless Hubble parameter in the phenomenological DE model of Chevalier, Polarski and Linder \cite{Chevallier:2000qy,Linder:2002et}. Eq.~(\ref{cpl}) reduces to $E_\Lambda$ when the DE component is a cosmological constant $\Lambda$, i.e. $(w_0$, $w_a)=(-1$, $0)$. The interpolating formula \eqref{H_z} of \cite{Cardone:2010} means that the expansion rate $H(z)$ for the two classes of $f(R)$ models we are considering may be obtained by interpolating the CPL and the \lcdm\ models back in time up to $z_\Lambda$, whilst it becomes exactly the same as in  \lcdm\ at earlier times $z>z_\Lambda$. In this \lcdm-like era, the matter density of the model is an effective value given by
\begin{equation}
\Omega_m^{(\mathrm{eff})}=\left\{\begin{array}{lr}
\frac{6c_2}{c_1+6c_2}&\mathrm{(HS)}\\
1-\frac{\lambda R_\star}{6\ho^2}&(\mathrm{St})
\end{array}\right..
\end{equation}

\subsection{Evolution of Cosmological Perturbations}
In the linear theory of cosmological perturbations and in the Newtonian gauge, the metric (\ref{flatFLRW}) takes the form
\begin{equation}
g_{\mu\nu}\,\de x^\mu\de x^\nu=-\left(1+2\Phi\right)\de t^2+\left(1+2\Psi\right)\de\mathbf x^2\label{pertFLRW},
\end{equation}
with
\begin{equation}
g_{\mu\nu}\equiv\overline g_{\mu\nu}+\delta g_{\mu\nu},
\end{equation}
where $\overline g_{\mu\nu}$ is the background metric and $\delta g_{\mu\nu}$ the perturbation. The two scalar potentials $\Phi$ and $\Psi$ are the metric perturbations in the Newtonian gauge. In this work we assume no anisotropic stress, thus the stress-energy tensor becomes
\begin{equation}
T_{\mu\nu}\equiv\overline T_{\mu\nu}+\delta T_{\mu\nu}
\end{equation}
with $\delta T^0_0=-\rho\delta$, $\delta T^0_i=-\rho v_{,i}$ and $\delta T^i_j=0$, in the matter dominated epoch. Here, $\delta=\delta\rho/\rho$ is the density contrast, $\mathbf v$ is the velocity of the scalar perturbations and a comma denotes a derivative with respect to the spatial coordinates.

Tsujikawa \cite{Tsujikawa:2007gd} showed that the evolution equation for the density contrast in $f(R)$ gravity models is
\begin{equation}
\ddot\delta+2H\dot\delta-4\pi\mathscr G(k,a)\rho\delta\simeq0,\label{eq:delta}
\end{equation}
where there is an effective gravitational constant
\begin{equation}
\mathscr G(k,a)=\frac{G}{f_{,R}}\frac{1+4\frac{k^2}{a^2}\frac{f_{,RR}}{f_{,R}}}{1+3\frac{k^2}{a^2}\frac{f_{,RR}}{f_{,R}}}.\label{G_Phi}
\end{equation}
The Poisson equation, which relates the potential $\Phi$ to the distribution of matter overdensities, in Fourier space reads
\begin{equation}
\Phi_k(a)=-4\pi\mathscr G(k,a)\frac{a^2}{k^2}\rho\delta_k(a).
\end{equation}
Finally, the other metric potential, $\Psi$, is related to the above quantities by the parameter $\eta\equiv-(\Phi+\Psi)/\Phi$, which characterises the strength of the effective anisotropic stress. In the present case, it is given by
\begin{equation}
\eta(k,a)=\frac{2\frac{k^2}{a^2}\frac{f_{,RR}}{f_{,R}}}{1+2\frac{k^2}{a^2}\frac{f_{,RR}}{f_{,R}}}.
\end{equation}
We remind that in standard GR, $\mathscr G(k,a)=G$ and $\eta=0$.

The power spectrum of the matter fluctuations $P^\delta(k,z)$ is obtained by taking the Fourier transform of the two-point correlation function of the density contrast solution of Eq.~\eqref{eq:delta}. This means
\begin{equation}
\langle \delta_k(z){\delta_{k'}}^\ast(z)\rangle=\left(2\pi\right)^3\delta_D\left(\mathbf k-\mathbf k'\right)P^\delta(k,z),
\end{equation}
where $\delta_D$ is the Dirac $\delta$ and $k\equiv|\mathbf k|$. Thus, the matter power spectrum can be written as
\begin{equation}
P^\delta(k,z)=2\pi^2{\delta_H}^2\left(\frac{k}{\ho^3}\right)^{n_s}T^2(k)\left[\frac{\delta_k(z)}{\delta_k(z=0)}\right]^2,
\end{equation}
with $n_s$ the tilt of the primordial power spectrum, $\delta_H$ its normalisation and $T(k)$ the matter transfer function, which describes the evolution of perturbations through the epochs of horizon crossing and radiation-matter transition.

In the non-linear r\'egime of the growth of matter overdensities, Eq.~\eqref{eq:delta} does not hold any more. In the \lcdm\ model, to obtain the non-linear matter power spectrum we need to resort to numerical solutions. A short cut is to use semi-analytical calculations \cite{Pietroni:2008jx} or fitting formul\ae\ extrapolated from numerical simulations (e.g. the \textsc{halofit} approach of \cite{Smith:2002dz}). For modified gravity to agree with Solar system observations, the non-linear matter power spectrum has to approach the standard $\Lambda$CDM solution on small scales. This means that the non-linear power spectrum has to be an interpolation of two power spectra. The former is the modified gravity non-linear power spectrum $P^\delta_\mathrm{MG}(k,z)$, which is obtained without the non-linear interactions that are responsible for the recovery of GR. This is equivalent to assume that gravity is modified down to small scales in the same way as in the linear r\'egime. The latter term, $P^\delta_\mathrm{GR}(k,z)$, is the non-linear power spectrum obtained in the DE model that follows the same expansion history of the Universe as the modified gravity model, yet obeying to GR. In other words, this is the non-linear power spectrum which will have a $\Lambda$CDM model with an expansions history $H(z)$ equivalent to that of the modified gravity theory.

Here, for the interpolation of these two spectra, we use the fitting formula proposed in Ref.~\cite{Hu:2007pj}
\begin{equation}
P^\delta(k,z)=\frac{P^\delta_\mathrm{MG}(k,z)+c_\mathrm{nl}(z)\Sigma^2(k,z)P^\delta_\mathrm{GR}(k,z)}{1+c_\mathrm{nl}(z)\Sigma^2(k,z)},
\end{equation}
where $\Sigma^2(k,z)=\left[k^3P^\delta_\mathrm{lin.}(k,z)/2\pi^2\right]^{a_1}$ picks out non-linear scales, since $P^\delta_\mathrm{lin.}(k,z)$ is the $f(R)$ linear power spectrum, and $c_\mathrm{nl}(z)=A\left(1+z\right)^{a_2}$ determines the scale at which the power spectrum approaches the GR result as a function of redshift. Their functional forms have been obtained by perturbation theory \cite{Koyama:2009me} and confirmed by $N$-body simulations \cite{Oyaizu:2008tb}. Here, we use $a_1=1/3$, $a_2=1.05$ and $A=0.08$.

\section{Cosmic Shear Tomography}\label{sec:wl}
In $f(R)$ models, for the two potentials $|\Phi|\neq|\Psi|$ holds, even though there is no formal anisotropic stress in the energy momentum tensor $T_{\mu\nu}$. Thus, the relation between the distribution of matter overdensities in the Universe and the two metric perturbations, the potentials $\Phi$ and $\Psi$, is not trivial. In GR, in the matter-dominated era, when there is no anisotropic stress, $\Phi=-\Psi$ and therefore we can simply use the Newtonian potential $\Phi$, thanks to the canonical Poisson equation, to compute cosmic convergence and shear. However, in general, the weak lensing effect is due to the combination of both the Newtonian and the metric potential. We will refer to this combination as the ``deflecting potential,'' and we will denote it with\footnote{In the literature, the combinations of the two potentials are often indicated by $\Phi_\pm=-\left(\Phi\pm\Psi\right)/2$. However, different authors use to refer to the metric potentials in the length element differently. For the sake of simplicity, we have chosen this notation, because we are only interested in the combined effect responsible for the weak lensing signal.}
\begin{equation}
\Upsilon\equiv-\frac{\Phi-\Psi}{2}.
\end{equation}

Weak gravitational lensing is responsible for the shearing and magnification of the images of high-redshift sources due to the presence of intervening matter. The distortion are due to fluctuations in the gravitational potential, and are directly related to the distribution of matter and to the geometry and dynamics of the Universe. In particular, the distortions occurring to a background image can be decomposed into a convergence $\kappa$ and a (complex) shear $\gamma=\gamma_1+i\gamma_2$, which are the entries of the distortion matrix
\begin{equation}
\mathscr D\equiv\left(\begin{array}{cc}
\kappa+\gamma_1&\gamma_2\\
\gamma_2&\kappa -\gamma_1
\end{array}\right).
\end{equation}
The distortion matrix is directly related to background and perturbed cosmological quantities, since
\begin{equation}
\mathscr D_{ij}=\int_0^\chi\!\!\de\chi'\,\chi'W(\chi')\Upsilon_{,ij}(\hat{\mathbf n},\chi')\label{phi,ij}
\end{equation}
\cite{Kaiser:1996tp,Bartelmann:1999yn}, and commas denote derivatives with respect to directions perpendicular to the line of sight. Here, $\de\chi=\de z/H(z)$ is the differential radial comoving distance,
\begin{equation}
W(\chi)=-2\chi\int_\chi^\infty\de\chi'\,\frac{\chi'-\chi}{\chi'}n(\chi')\label{W(z)}
\end{equation}
is the weight function of weak lensing, and $n\left[\chi(z)\right]$ represents the redshift distribution of the sources, such that $\int\!\!\de\chi\,n(\chi)=1$.

In the flat-sky approximation, we expand the shear $\gamma(\hat{\mathbf n})$ in its Fourier modes
\begin{equation}
\gamma(\hat{\mathbf n})=\int\!\!\frac{\de^2\ell}{{(2\pi)}^2}\,\gamma(\boldsymbol{\ell})e^{i\boldsymbol{\ell}\cdot\hat{\mathbf n}}.
\end{equation}
The power spectrum is defined as the Fourier transform of the 2D correlation function
\begin{equation}
\langle\gamma(\boldsymbol{\ell})\gamma^\ast(\boldsymbol{\ell}')\rangle={(2\pi)}^2\delta_D(\boldsymbol{\ell}-\boldsymbol{\ell}')C^\gamma(\ell).
\end{equation}
Thus, we have \cite{Kaiser:1991qi}
\begin{equation}
C^\gamma(\ell)=\frac{\ell^4}{4}\int\!\!\de\chi\,\frac{W^2(\chi)}{\chi^6}P^\Upsilon\left(\frac{\ell}{\chi},\chi\right),\label{C(l)}
\end{equation}
where $P^\Upsilon(k,z)$ is the power spectrum of the deflecting potential, and we have introduced Limber's approximation, where the only Fourier modes that contribute to the integral are those with $\ell=k\chi$.

In the case where one has distance information for individual sources, we can use this information for statistical studies. A natural course of action is to divide the survey into slices at different distances, and perform a study of the shear pattern on each slice \cite{Hu:1999ek}. In order to use the information effectively, it is necessary to look at cross-correlations of the shear fields in the slices, as well as correlations within each slice. This procedure is usually referred to as tomography. To better constrain these $f(R)$ models with cosmic shear, we perform cosmic shear tomography. To do so, we separate the redshift distribution of sources $n[\chi(z)]$ into redshift bins, each roughly containing the same number of sources. By doing so, the 2-dimensional shear power spectrum in the flat-sky approximation \eqref{C(l)} reads
\begin{equation}
C^\gamma_{ij}(\ell)=\frac{\ell^4}{4}\int\!\!\de\chi\,\frac{W_i(\chi)W_j(\chi)}{\chi^6}P^\Upsilon\left(\frac{\ell}{\chi},\chi\right),\label{eq:tomography}
\end{equation}
where $W_i(\chi)$ is the weight function \eqref{W(z)} related to the $i$th bin.

\section{Fisher Matrix Analysis}\label{sec:fisher}
Cosmological parameters influence the shear in a number of ways, for instance the matter power spectrum $P^\delta(k,z)$ depends on $\om$, $h$ and the linear amplitude $\sigma_8$. The linear power spectrum depends on the growth rate, which is also sensitive to the parameter of the $\Lambda$-like equation of state $w_\Lambda=p_\Lambda/\rho_\Lambda$. It also affects the $\chi(z)$ relation and hence the angular diameter distance $\sin_K\left[\chi(z)\right]$. These parameters $\{\vartheta_\alpha\}$ may be estimated from the data using likelihood methods. Assuming uniform priors for the parameters, the maximum a posteriori probability for the parameters is given by the maximum likelihood solution. We use a Gaussian likelihood
\begin{equation}
2\ln L=-\mathrm{Tr}\left[\ln\mathcal C+\mathcal C^{-1}\mathcal D\right],
\end{equation}
where $\mathcal C=\langle(\mathbf d-\mathbf d^\mathrm{th})(\mathbf d-\mathbf d^\mathrm{th})^T\rangle$ is the covariance matrix and $\mathcal D=(\mathbf d-\mathbf d^\mathrm{th})(\mathbf d-\mathbf d^\mathrm{th})^T$ is the data matrix, with $\mathbf d$ the data vector and $\mathbf d^\mathrm{th}$ the theoretical mean vector.

The expected errors on the parameters can be estimated with the Fisher information matrix \cite{Fisher:1935,Jungman:1995bz,Tegmark:1996bz}. This has the advantage that different observational strategies can be analysed and this can be very valuable for experimental design. The Fisher matrix gives the best errors to expect, and should be accurate if the likelihood surface near the peak is adequately approximated by a multivariate Gaussian.

The Fisher matrix is the expectation value of the second derivative of $\ln L$ with respect to the parameters $\{\vartheta_\alpha\}$, i.e.
\begin{equation}
\mathcal F_{\alpha\beta}=-\left\langle\frac{\partial^2\ln L}{\partial\vartheta_\alpha\partial\vartheta_\beta}\right\rangle\label{fisherm}
\end{equation}
and the marginal error on parameter $\vartheta_\alpha$ is $\left[\left(\mathcal F^{-1}\right)_{\alpha\alpha}\right]^{\frac{1}{2}}$. If the means of the data are fixed, the Fisher matrix can be calculated from the covariance matrix and its derivatives \cite{Tegmark:1996bz} by
\begin{equation}
\mathcal F_{\alpha\beta}=\frac{1}{2}\mathrm{Tr}\left[\mathcal C^{-1}\mathcal C_{,\alpha}\mathcal C^{-1}\mathcal C_{,\beta}\right].
\end{equation}
For a square patch of the sky, the Fourier transform leads to uncorrelated modes, provided the modes are separated by $2\pi/\Theta_\mathrm{rad}$ where $\Theta_\mathrm{rad}$ is the side of the square in radians, and the Fisher matrix is simply the sum of the Fisher matrices of each $\ell$ mode,
\begin{equation}
\mathcal F_{\alpha\beta}=\sum_\ell\frac{2\ell+1}{2}f_\mathrm{sky}\mathrm{Tr}\left[\left(\mathcal C^\ell\right)^{-1}{\mathcal C^\ell}_{,\alpha}\left(\mathcal C^\ell\right)^{-1}{\mathcal C^\ell}_{,\beta}\right],\label{eq:Fisher}
\end{equation}
where $f_\mathrm{sky}$ is the fraction of the sky covered by the survey under analysis, and $\mathcal C^\ell$ is the covariance matrix for a given $\ell$ mode.

\subsection{Bayesian Model Selection}\label{ssec:B-evidence}
In this paper we compute parameter forecasts from cosmic shear tomography for viable $f(R)$ models. It is worth noticing that we are dealing with a model alternative to standard $\Lambda$CDM. Hence, besides determining the best-fit value -- and the errors -- on a set of parameters within a model, we can also ask if this particular alternative model is preferable to the standard. Model selection is in a sense a higher-level question than parameter estimation. While in estimating parameters one assumes a theoretical model within which one interprets the data, in model selection one wants to know which theoretical framework is preferred given the data. Clearly, if our alternative model has more parameters than the standard one, chi-square analysis will not be of any use, because it will always reduce if we add more degrees of freedom. Bayesian analysis provides a useful ``Occam's razor,'' which involves computation of the Bayesian evidence and of the Bayes factor $B$.

The Bayesian evidence of a model $\mathcal M$ is defined as the marginalisation over the parameters
\begin{equation}
p(\mathbf d|\mathcal M)=\int\!\!\de^m\vartheta\,p(\mathbf d|\boldsymbol{\vartheta},\mathcal M)p(\boldsymbol{\vartheta}|\mathcal M),
\end{equation}
where $\boldsymbol{\vartheta}$ is the parameter vector, and $p(\mathbf d|\boldsymbol{\vartheta},\mathcal M)$ is the marginal likelihood %$L(\boldsymbol{\vartheta},\mathcal M;\mathbf d)$ 
of the parameters $\boldsymbol{\vartheta}$ of the model $\mathcal M$ given the data $\mathbf d$. Let us consider two competing models $\mathcal M_1$ and $\mathcal M_2$, the former nested in the latter. This means that $\mathcal M_1$ is simpler, because the set of its parameters $\{\vartheta_{\alpha_1}\}$ is contained in the $\mathcal M_2$ parameter set $\{\vartheta_{\alpha_2}\}$, with $\alpha_1=1,\ldots,n_1$ and $\alpha_2=1,\ldots,n_2$, $n_2>n_1$. In such a situation, one can compute the Bayes factor $B$, which is the ratio of the two corresponding posterior evidence probabilities $p(\mathcal M_1|\mathbf d)$ and $p(\mathcal M_2|\mathbf d)$. The posterior probability for each model $\mathcal M_i$ is given by Bayes' theorem
\begin{equation}
p(\mathcal M_i|\mathbf d)=\frac{p(\mathbf d|\mathcal M_i)p(\mathcal M_i)}{p(\mathbf d)}.
\end{equation}

If we choose noncommittal, constant priors $p(\mathcal M_1)=p(\mathcal M_2)=1/2$, the ratio of the posterior evidence probabilities reduces to the ratio of the evidence. Heavens et al. \cite{Heavens:2007ka} showed that, in the Laplace approximation, where the expected likelihoods are given by multivariate Gaussians, and if one considers $\langle B\rangle$ as the ratio of the expected values, rather than the expectation value of the ratio, one eventually gets
\begin{equation}
\langle B\rangle=\frac{\sqrt{\det\mathcal F_2}}{\sqrt{\det\mathcal F_1}}\left(2\pi\right)^{-\frac{p}{2}}\prod_{q=1}^p\Delta\vartheta_{\alpha_1+q}e^{-\frac{1}{2}\boldsymbol{\delta\vartheta}\mathcal F_2\boldsymbol{\delta\vartheta}}.\label{B}
\end{equation}
Here, $\mathcal F_i$ is the Fisher matrix relative to the $i$th model, $p=n_2-n_1$ is the number of extra parameters, and $\boldsymbol{\delta\vartheta}$ is the vector of the parameter shifts. Indeed, if the correct underlying model is $\mathcal M_2$, the maximum of the expected likelihood will not, in general, be at the correct parameter values of $\mathcal M_1$ (see Fig.~1 of \cite{Heavens:2007ka}). The $n_1$ parameters of $\mathcal M_1$ shift their values to compensate the fact that $\vartheta_{\alpha_1+1},\ldots,\vartheta_{\alpha_1+p}$ are kept fixed at an incorrect fiducial value $\vartheta_{\alpha_1+1}=\ldots=\vartheta_{\alpha_1+p}=0$. These shifts can be computed  under the assumption of a multivariate Gaussian distribution \cite{Taylor:2006aw}, and read
\begin{equation}
\boldsymbol{\delta\vartheta}=-{\mathcal F_1}^{-1}\mathcal G_2\boldsymbol{\delta\psi},\label{shifts}
\end{equation}
with $\mathcal G_2$ a subset of the $\mathcal M_2$ Fisher matrix and $\boldsymbol{\delta\psi}$ the shifts of the $p$ extra parameters $\boldsymbol\psi$.

It is usual to consider the logarithm of the Bayes factor, for which the so-called ``Jeffreys' scale'' gives empirically calibrated levels of significance for the strength of evidence \cite{Jeffreys:1961}. A more recent version of Jeffreys' scale sets $1<|\ln B|<2.5$ as ``substantial'' evidence in favour of a model, $2.5<|\ln B|<5$ as ``strong,'' and $|\ln B|>5$ as ``decisive.'' These descriptions seem too aggressive, for $|\ln B|=1$ corresponds to a posterior probability for the less-favoured model which is $0.37$ times the favoured model \cite{Kass95bayesfactors}. Other authors have introduced different terminology (e.g. \cite{Trotta:2005ar}).

\section{Results}\label{sec:results}
We compute our results for a flat Universe with cosmological parameters $h=0.7$, $\om=0.28$, $\ob=2.22\cdot10^{-2}h^{-2}$, $\ol=1-\om$, $\log_{10}\lambda/\log_{10}c_1=2.38/4.98$, $\log_{10}\kappa/\log_{10}c_2=-2.6/3.79$ and $\nst/\nhs=1.79/1.64$, where $\kappa=R_\star/R_0$, as obtained with the Monte Carlo Markov Chains of Ref.~\cite{Cardone:2010}. For the matter power spectrum, we use the transfer function proposed by Eisenstein \& Hu (1998) \cite{Eisenstein:1997ik}, rms mass fluctuations $\sigma_8=0.8$ on a scale of $8\,h^{-1}\,\mathrm{Mpc}$, and spectral index $n_s=0.96$ \cite{Larson:2010gs}.

\begin{figure}[ht!]
\centering
\includegraphics[width=0.9\textwidth]{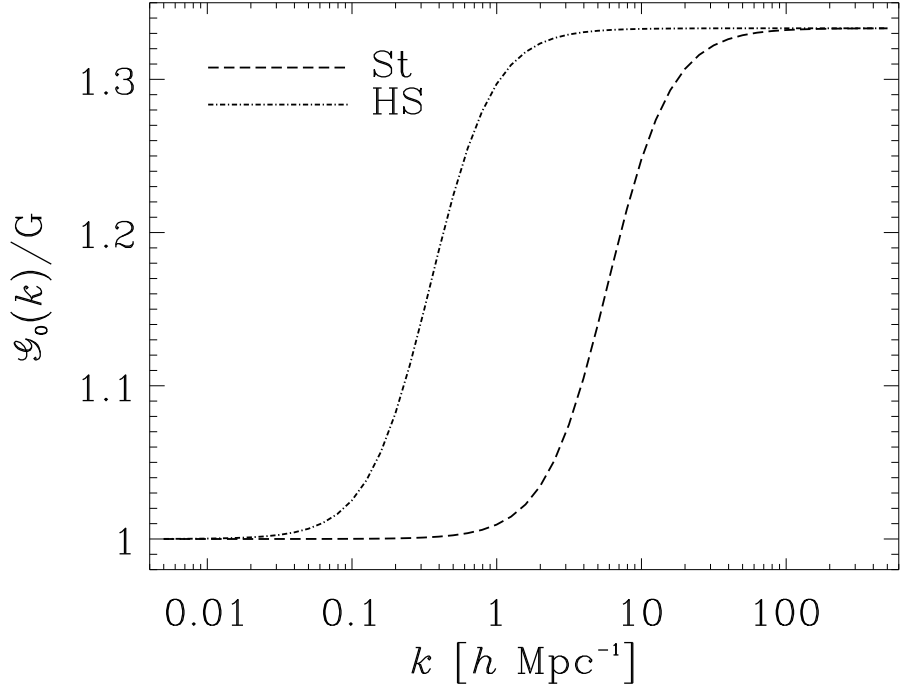}
\caption{The present-day effective gravitational constant $\mathscr G_0(k)\equiv\mathscr G(k,a=a_0)$ versus the physical scale $k$, normalised to the Newtonian constant $G$.}\label{fig:Geff}
\end{figure}

In $f(R)$ models, gravity is stronger than in GR \cite{Tsujikawa:2007gd}. In particular, the effective gravitational constant $\mathscr G(k,a)$, which appears in the source term driving the evolution of matter density perturbations (Eq.~\ref{eq:delta}), can change significantly compared to the Newtonian gravitational constant $G$. It has been shown that in the so-called ``scalar-tensor'' r\'egime $\mathscr G\sim4G/3$ \cite{Tsujikawa:2007gd}. In Fig.~\ref{fig:Geff} we show the present-day value $\mathscr G(k,a=a_0)$ of the effective gravitational constant normalised to the Newtonian $G$. At large scales, $f(R)$ gravity behaves like GR, whilst the larger is the value of $k$, the greater is the difference. In more detail, the HS model presents a gravitational coupling larger than the St model, because it reaches the $4G/3\simeq1.33G$ value at smaller scales.

\subsection{The Cosmic Shear Signal}
We compute our results for a $20,000\,\mathrm{deg}^2$ next-generation cosmic shear experiment such as Euclid.\footnote{http://sci.esa.int/science-e/www/area/index.cfm?fareaid=102} We use the Euclid baseline given in the Yellow Book \cite{Cimatti:2009is}. The source distribution over redshifts has the form \cite{1994MNRAS.270..245S}
\begin{equation}
n(z)\propto z^2e^{-\left(\frac{z}{z_0}\right)^{1.5}},\label{eq:n_z-Euclid}
\end{equation}
where $z_0=z_m/1.4$, and $z_m=0.89$ is the median redshift of the survey. The number density of the sources, with estimated redshift and shape, is $35$ per square arcminute. We also compute the expected errors according to \cite{Kaiser:1991qi,Kaiser:1996tp}
\begin{equation}
\Delta C^\gamma(\ell)=\sqrt{\frac{2}{(2\ell+1)f_\mathrm{sky}}}\left[C^\gamma(\ell)+\frac{\langle{\gamma_\mathrm{int}}^2\rangle}{\bar n}\right],\label{eq:errorbars}
\end{equation}
where $f_\mathrm{sky}=\Theta_\mathrm{deg}^2\pi/129,600$ is the fraction of the sky covered by a survey of area $\Theta_\mathrm{deg}^2$ and $\langle{\gamma_\mathrm{int}}^2\rangle^{0.5}\simeq0.4$ is the galaxy-intrinsic shear rms in one component.

\begin{figure}[!h]
\includegraphics[width=0.9\textwidth]{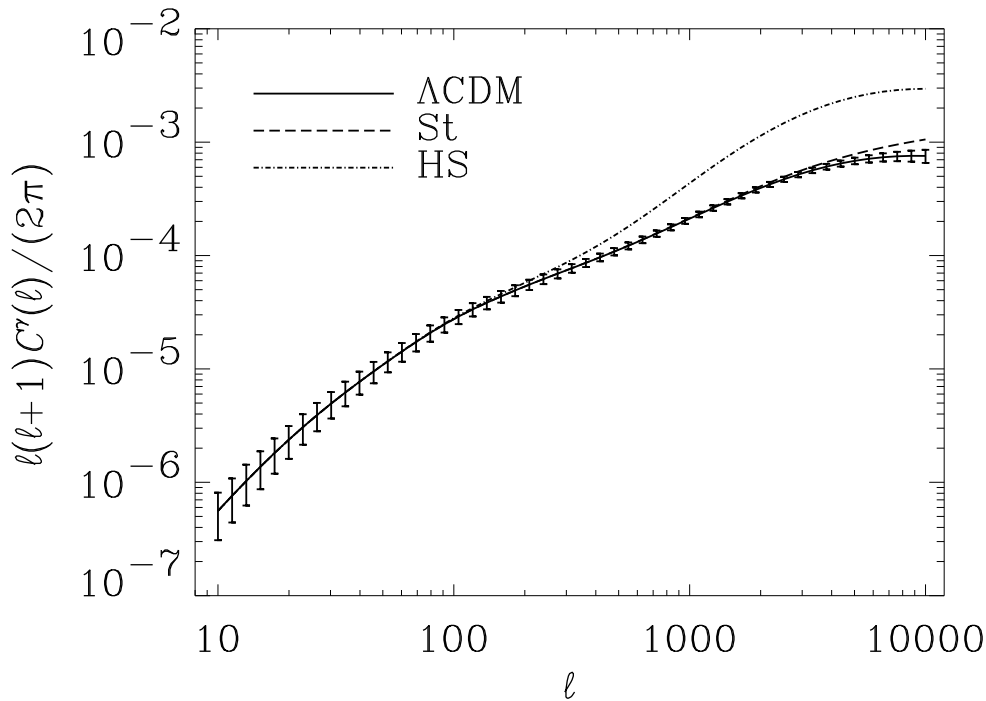}
\caption{Shear power spectrum $\ell(\ell+1)C^\gamma(\ell)/(2\pi)$ in the \lcdm\ (solid), St (dashed) and HS (dot-dashed) models; $1\sigma$ errorbars on the \lcdm\ signal, computed according to Eq.~\eqref{eq:errorbars}, are also shown.}\label{fig:Euclid}
\end{figure}

The peculiarities of the two $f(R)$ models depicted in Fig.~\ref{fig:Geff} propagate into the cosmic shear power spectrum $C^\gamma(\ell)$, which is shown in Fig.~\ref{fig:Euclid}. The curves refer to the standard \lcdm\ cosmology (solid), the St (dashed) and the HS (dot-dashed) models, respectively. At large angular scales the three signals are indistinguishable, while for large values of $\ell$, differences start to be important. This is a consequence of Limber's approximation, which sets $k=\ell/\chi$ and therefore shows at large $\ell$'s the features $\mathscr G_0(k)$ presents at large $k$'s. Thus, we observe in the cosmic shear power spectrum the same behaviour of the effective gravitational constant $\mathscr G(k,a)$. Specifically, the HS signal differs from \lcdm\ by more than $1\sigma$ at $\ell\ge 300$. On the contrary, the St model produces a cosmic shear power spectrum in agreement with \lcdm\ up to $\ell\simeq5000$, and the errorbars show that it is almost completely degenerate with \lcdm.

\begin{figure}
\centering
\includegraphics[width=\textwidth]{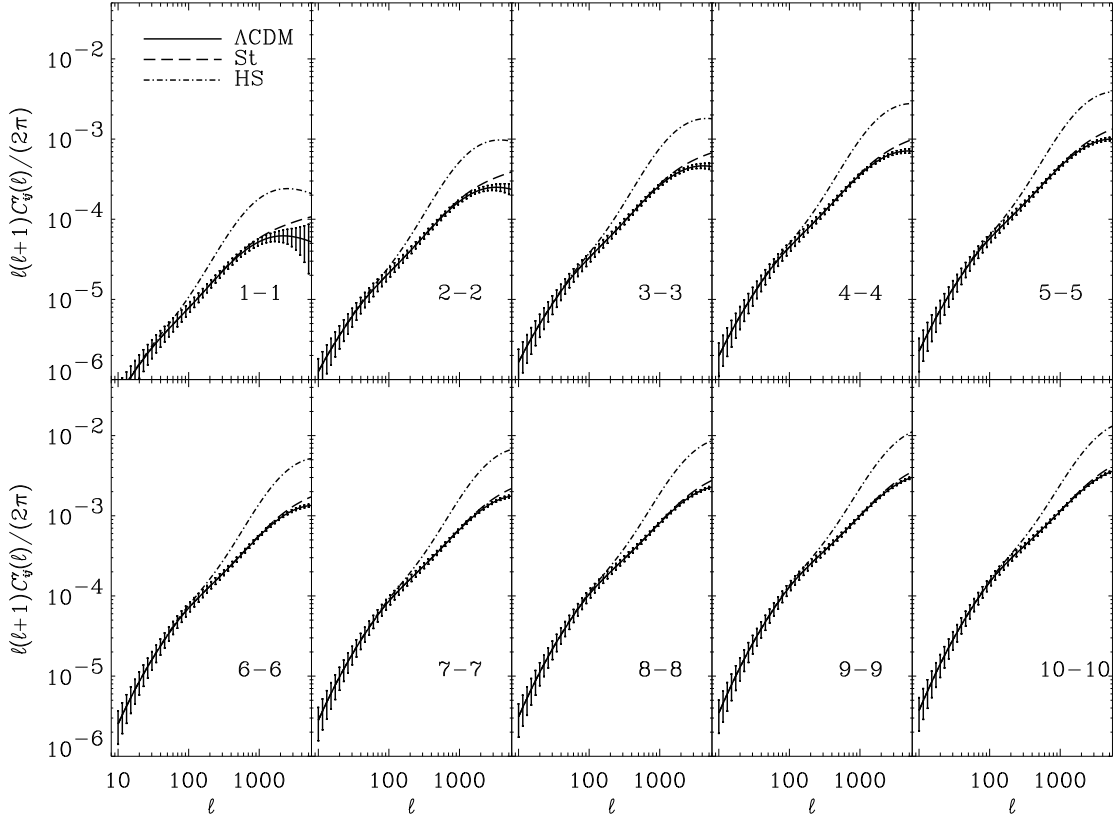}
\caption{Diagonal elements $C^\gamma_{ii}(\ell)$ of the tomographic shear matrix for the \lcdm\ (solid), St (dashed) and HS (dot-dashed) models. Curves refer to shear autocorrelations, from the $1^\mathrm{st}$-$1^\mathrm{st}$ to the $10^\mathrm{th}$-$10^\mathrm{th}$ bin pairs with increasing redshift; $1\sigma$ errorbars on the \lcdm\ signal, computed according to Eq.~\eqref{eq:errorbars}, are also shown.}\label{fig:Euclid-binning}
\end{figure}

To better constrain the model parameters, and with the aim of lifting the degeneracy between \lcdm\ and the St model, we calculate the tomographic shear matrix $C^\gamma_{ij}(\ell)$. To do so, we separate the Euclid distribution of sources \eqref{eq:n_z-Euclid} into ten redshift bins. Fig.~\ref{fig:Euclid-binning} presents the diagonal elements $C^\gamma_{ii}(\ell)$ of the tomographic shear matrix as a function of the angular scale $\ell$ for standard \lcdm\ cosmology (solid), the St (dashed) and the HS (dot-dashed) models, respectively. We also show $1\sigma$ errorbars on the \lcdm\ signal. Though tomography does enhance the quality of the signal when one looks at the high-redshift autocorrelations, we find that the degeneracy between \lcdm\ and the St model is not removed. The off-diagonal elements of the tomographic shear matrix show a similar behaviour. In principle, they should be more useful, since the Poissonian noise term in Eq.~\eqref{eq:errorbars} holds for correlations between the same bin only. However, the St signal is still too close to what predicted by \lcdm, at least for the range of angular scales probed here.

\subsection{Cosmological Parameter Forecasts}
Once we have introduced the method and the survey design formalism, we now show the cosmological parameter forecasts for such a survey. By using the Fisher matrix analysis outlined in Ref.~\cite{Taylor:2006aw}, we calculate the predicted Fisher matrices and the parameter constraints for a $20,000\,\mathrm{deg}^2$ Euclid-like survey. In all Fisher matrix calculations we use a six-parameter cosmological set $\{\om=\odm+\ob$, $\ob$, $h$, $\log_{10}\lambda/\log_{10}c_1$, $\log_{10}\kappa/\log_{10}c_2$, $\nst/\nhs\}$. Note that we do not use $\ol$ as a free parameter.

It is worth noting that, in the high-wavenumber r\'egime, the fitting formul\ae\ of \cite{Smith:2002dz} may be unreliable, or baryonic effects might alter the power spectrum ($k>10\,h\,\mathrm{Mpc}^{-1}$ \cite{White:2004kv,Zhan:2004wq}). Hence, we do not analyse modes with $k>1.5\,h\,\mathrm{Mpc}^{-1}$. However, in Limber's approximation $\ell\propto k^{-1}$, the choice of the maximum angular wavenumber $\ell_\mathrm{max}$ entering Eq.~\eqref{eq:Fisher} is crucial. The largest proper $\ell_\mathrm{max}$ is a matter of debate (e.g. \cite{vanDaalen:2011xb,Semboloni:2011fe}). Here, we follow three approaches, one more conservative with $\ell_\mathrm{max}=1000$, one optimistic with $\ell_\mathrm{max}=5000$ -- which was already used in testing alternative cosmological models with cosmic shear \cite{Camera:2010wm} --, and one where $\ell_\mathrm{max}$ is determined in a bin-dependent way. Actually, fixing the distances of the tomographic binning removes some flexibility in the probed physical wavenumbers; so there is a risk that some useful modes are excluded (thus increasing the statistical errors), and/or that, for the nearby shells, the sampled physical wavenumber range extends to too high a value of $k$, where theoretical uncertainties become a potential source of systematic error. To avoid these potential problems, we define $\ell_\mathrm{max}=\chi(z_\mathrm{bin})/k_\mathrm{max}$ \cite{Kitching:2010wa}: we thus increase the largest angular scale for distant shells and reduce it for nearby shells.

\begin{figure}[ht!]
\centering
\includegraphics[width=0.9\textwidth]{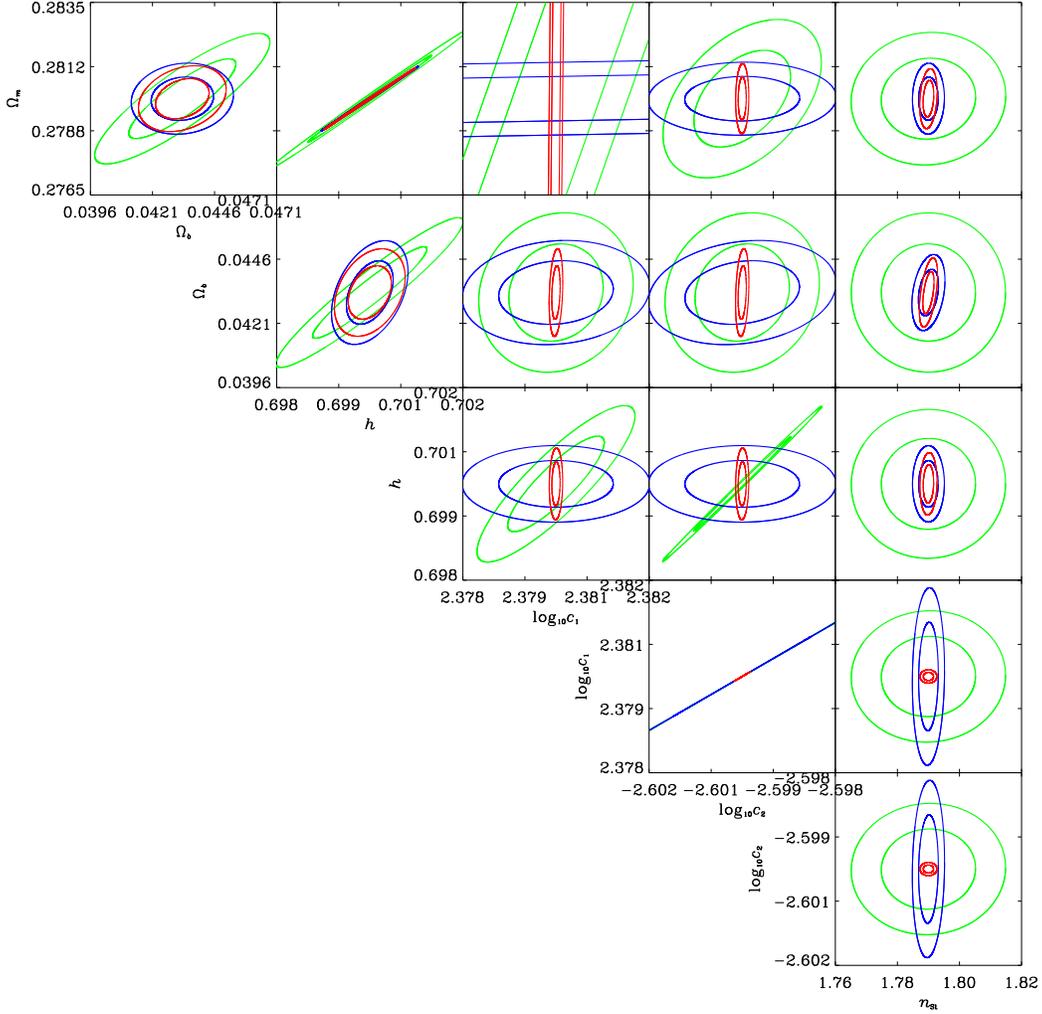}
\caption{Expected marginal errors on the St model cosmological parameters from a $20,000\,\mathrm{deg}^2$ Euclid-like survey. Ellipses show the $1\sigma$ and $2\sigma$ errors for two parameters ($68.3\%$ and $95.8\%$ confidence regions, respectively), marginalised over all the other parameters. The green, red and blue curves refer to the conservative, bin-dependent and optimistic $\ell_\mathrm{max}$.}\label{fig:ellipses-St}
\end{figure}

\begin{figure}[ht!]
\centering
\includegraphics[width=0.9\textwidth]{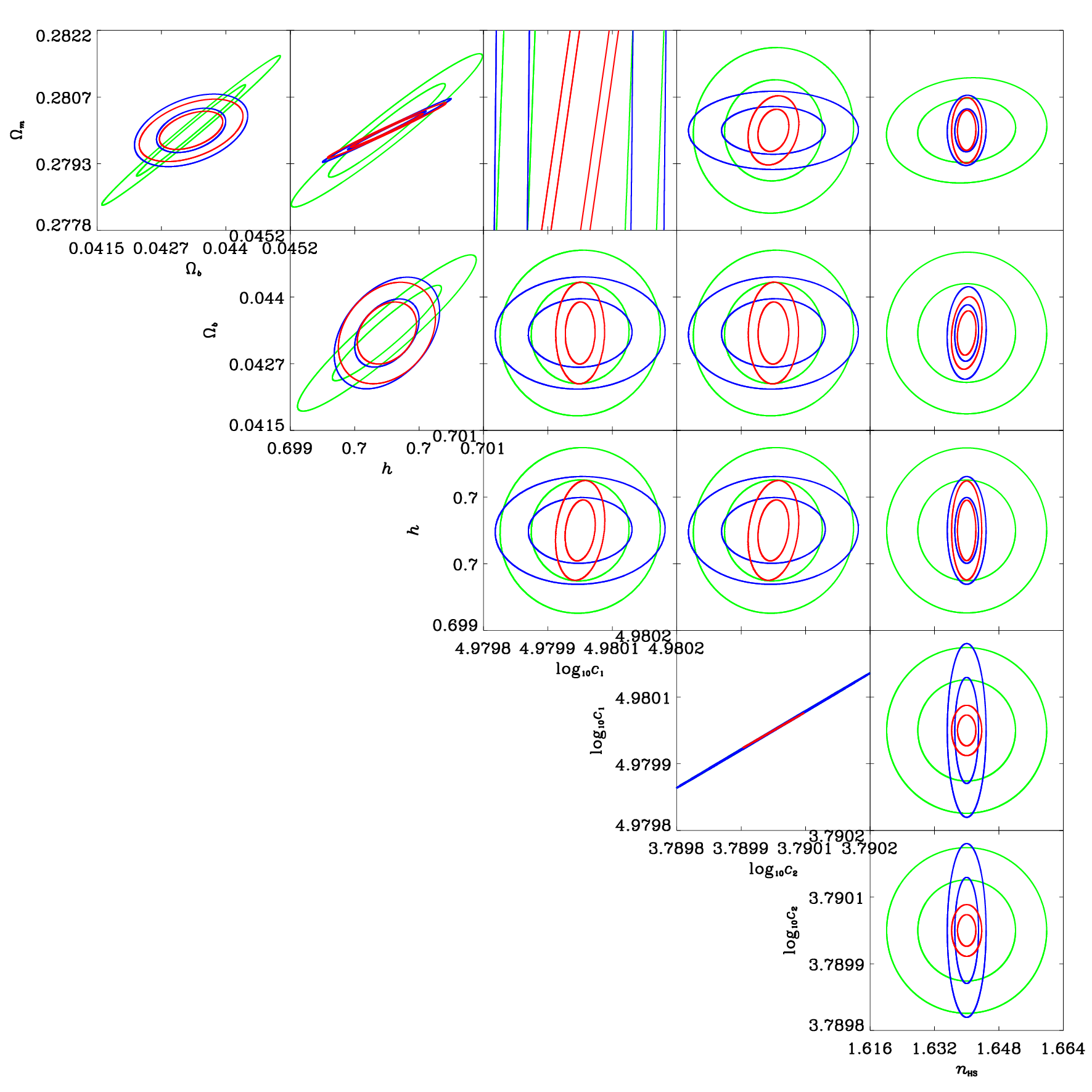}
\caption{Same as Fig.~\ref{fig:ellipses-St} for the HS model parameters.}\label{fig:ellipses-HS}
\end{figure}

Figs.~\ref{fig:ellipses-St}-\ref{fig:ellipses-HS} show the $1\sigma$ and $2\sigma$ contours obtained by calculating the Fisher matrices of the tomographic shear signal \eqref{eq:tomography}, for the St and HS models, respectively. The green curves refer to $\ell_\mathrm{max}=1000$, the blue ones to $\ell_\mathrm{max}(z_\mathrm{bin})$ and the red ones to $\ell_\mathrm{max}=5000$.

The use of tomography enhances the accuracy significantly, compared to the cosmic shear power spectrum. This analysis thus yields tighter constraints on the model parameters. Specifically, we show our constraints in Tables~1-2, for the St and HS models, respectively, where $\mu_\vartheta$ is the mean fiducial value of the parameter and $\sigma_\vartheta$ is the predicted standard deviation. However, these small constraints may probably under-predict the true parameter errors. This is due to the fact that gravitational lensing does mix angular modes to some degree because of its intrinsic non-local structure. Non-linear mode coupling affects Fisher matrices for cosmological observations. Indeed, these couplings tend to correlate small-scale power, moving information from lower to higher-order moments of the shear field. Therefore, Gaussian approximations may produce over-optimistic forecasts \cite{Kiessling:2011gv}.
\TABULAR[ht!]{cccccc}{\multicolumn{5}{c}{St}\\$\vartheta$ & $\mu_\vartheta$ & $\sigma_\vartheta^{\ell_\mathrm{max}=1000}$ & $\sigma_\vartheta^{\ell_\mathrm{max}=5000}$ & $\sigma_\vartheta^{\ell_\mathrm{max}(z_\mathrm{bin})}$ \\
\hline
$\om$ & $0.28$ & $0.0012$ & $0.0005$ & $0.0005$\\
$\ob$ & $2.22\cdot10^{-2}h^{-2}$ & $0.0012$ & $0.0008$ & $0.0007$\\
$h$ & $0.7$ & $0.0006$ & $0.0003$ & $0.0003$\\
$\log_{10}\lambda$ & $2.38$ & $0.00055$ & $0.00067$ & $0.00005$\\
$\log_{10}\kappa$ & $-2.6$ & $0.00059$ & $0.00072$ & $0.00005$\\
$\nst$ & $1.79$ & $0.0097$ & $0.0020$ & $0.0011$}{Forecasts on the St model parameters.}
\TABULAR[ht!]{cccccc}{\multicolumn{5}{c}{HS}\\$\vartheta$ & $\mu_\vartheta$ & $\sigma_\vartheta^{\ell_\mathrm{max}=1000}$ & $\sigma_\vartheta^{\ell_\mathrm{max}=5000}$ & $\sigma_\vartheta^{\ell_\mathrm{max}(z_\mathrm{bin})}$ \\
\hline
$\om$ & $0.28$ & $0.0007$ & $0.0003$ & $0.0003$\\
$\ob$ & $2.22\cdot10^{-2}h^{-2}$ & $0.0006$ & $0.0004$ & $0.0004$\\
$h$ & $0.7$ & $0.0004$ & $0.0003$ & $0.0002$\\
$\log_{10}c_1$ & $4.98$ & $0.00008$ & $0.00008$ & $0.00002$\\
$\log_{10}c_2$ & $3.79$ & $0.00008$ & $0.00008$ & $0.00002$\\
$\nst$ & $1.64$ & $0.0081$ & $0.0019$ & $0.0015$}{Same as Table~1 for the HS model.}

\subsection{Model Selection}\label{selection}
In Section~\ref{ssec:B-evidence} we showed how the Bayes factor can be used to determine which model is favoured by the data. We outlined how the issue of Bayesian model selection can be performed in the case of nested models. Actually, only the HS model is formally nested in \lcdm. Indeed, if one sets $\nhs=0$, the two extra parameters, $c_1$ and $c_2$, depend on each other. Specifically,
\begin{equation}
\log_{10}c_1=6\left(1+10^{\log_{10}c_2}\right)\frac{\ol}{\om}
\end{equation}
must hold.

\begin{figure}[!h]
\includegraphics[width=0.9\textwidth]{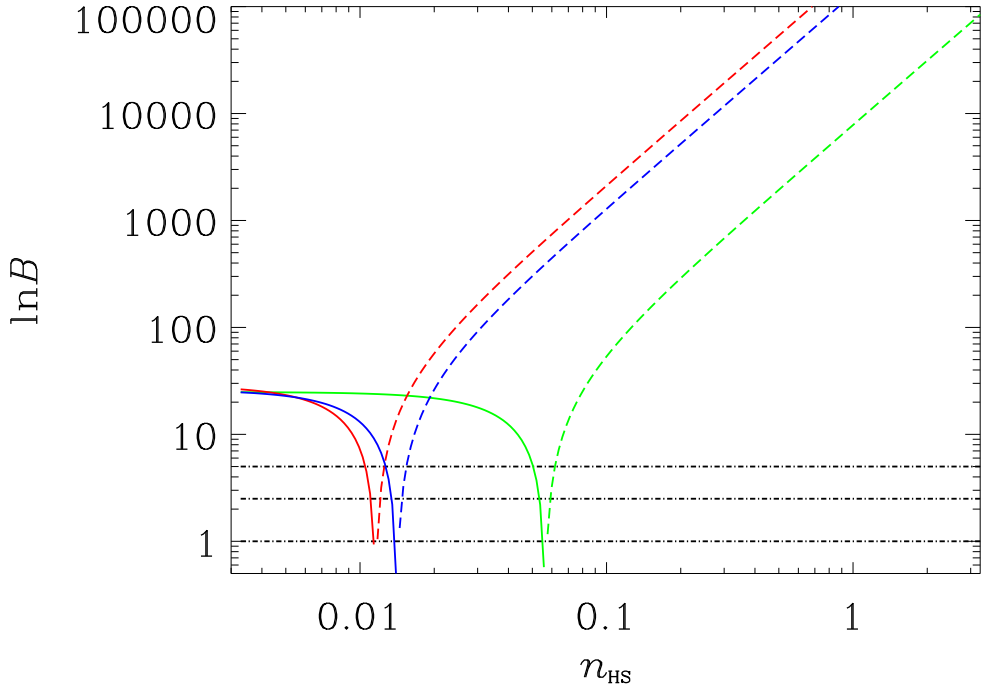}
\caption{The Bayes factor $\ln B$ for the HS model (nested case) over standard $\Lambda$CDM as a function of the extra parameter $\nhs$. The green, red and blue curves refer to the conservative, bin-dependent and optimistic $\ell_\mathrm{max}$, respectively. The horizontal lines denote the Jeffreys' scale levels of significance.}\label{fig:lnB}
\end{figure}

By using the Fisher matrix formalism for a Euclid-like survey, we compute the Bayes factor $\ln B$ as a function of $\nhs$ for the HS model over the standard $\Lambda$CDM cosmology. The result is presented in Fig.~\ref{fig:lnB}. The green, red and blue curves refer to $\ell_\mathrm{max}=1000,\,5000$ and $\ell_\mathrm{max}(z_\mathrm{bin})$, respectively. A solid line denotes a positive value of $\ln B$, whereas a dashed line means $\ln B<0$. Therefore, if Euclid detects any nonzero $\nhs$'s, the Bayes factor will favour the HS model only on the right of the cusps. Contrarily, Occam's razor will prefer \lcdm\ anyway. The three dot-dashed horizontal lines show the ``strength'' of this favour/disfavour, being them $\ln B=1,\,2.5$ and $5$, as in the current version of Jeffreys' scale. We want to emphasise that the HS model value $\nhs=1.64$ lies in the region of ``decisive'' evidence in favour of the more complex model. Thus, Euclid will favour the HS model unquestionably, if $\nhs=1.64$ will be measured.

\section{Conclusions}\label{sec:conclusions}
In this paper, we study whether the cosmic shear signal measured by the ESA Cosmic Vision Euclid satellite \cite{2008SPIE.7010E..38R,Cimatti:2009is,Refregier:2010ss} can constrain two viable $f(R)$ models. These models, proposed by Starobinsky \cite{Starobinsky:2007hu} (St) and Hu \& Sawicki \cite{Hu:2007nk} (HS) to be able to pass Solar System tests, account for the present-day accelerated expansion of the Universe without any cosmological constant term. At the same time, the functional form of their $f(R)$ correctly reproduces the Hubble rate of an early-time matter-dominated cosmos which undergoes a late-time phase of accelerated expansion. Recently, Thomas et al. \cite{Thomas:2011pj} performed a similar analysis, but they focused on constraining the scalaron mass and its parameterisations.

Cardone et al. \cite{Cardone:2010} derived the St and HS model parameters against several cosmological datasets by fitting the expansion rate history $H(z)$. Specifically, they used the Hubble diagram of SneIa \cite{Hicken:2009dk} and Gamma Ray Bursts \cite{Cardone:2010}. Moreover, they use $H(z)$ data from passively evolving red galaxies \cite{Freedman:2000cf}, Baryon Acoustic Oscillations extracted from the seventh data release of the Sloan Digital Sky Survey, and distance priors from the recent WMAP7 data \cite{Larson:2010gs}. Therefore, we use Cardone et al. \cite{Cardone:2010} values for the model parameters to compute our observables.

We calculate the cosmic shear power spectra and tomographic shear matrices for the two models and compared them with \lcdm. Beynon et al. \cite{Beynon:2009yd} showed that there is substantial additional discriminatory power between $f(R)$ models by the inclusion of the non-linear power r\'egime. They have also shown that using only the \textsc{halofit} formula \cite{Smith:2002dz} without any attempt to obtain the GR non-linear power spectrum on small scales leads to an overestimation in the ability of future surveys to differentiate between different growth histories. Similarly, Casarini et al. \cite{Casarini:2011ms} demonstrated that \textsc{halofit} expressions, well tested for the \lcdm\ model, implies substantial discrepancies with respect to results directly obtained from $N$-body simulations, when the effective $w_\mathrm{DE}(z)\neq-1$. To avoid all these problems, we use the fitting formul\ae\ proposed in Ref.~\cite{Hu:2007pj}, which are an interpolation of the modified-gravity non-linear matter power spectrum and the small-scales GR prediction. These formul\ae\ have been confirmed both theoretically \cite{Koyama:2009me} and by $N$-body simulations \cite{Oyaizu:2008tb}, although for a slightly different $f(R)$ model.

In $f(R)$ models, gravity is stronger than in GR \cite{Tsujikawa:2007gd}, and in particular the Newtonian constant of gravitation is replaced by a time- and scale-dependent effective $\mathscr G(k,a)$. Furthermore, when gravity is not described by GR, the two metric potentials $\Phi$ and $\Psi$ are no longer equal in modulus. Therefore, one has to take the deflecting potential $\Upsilon=-(\Phi-\Psi)/2$ into account, when computing lensing observables. As expected, we find that the cosmic shear signal tracks the differences between the effective gravitational constant in the St, HS and \lcdm\ models. Specifically, the HS shear signal is different from \lcdm\ at angular scales $\ell\gtrsim 300$, whilst for the St model the agreement holds for all $\ell$ up to $\ell\simeq5000$. We also find that the difference between the \lcdm\ cosmic shear signal and what is expected for the St model are too weak to allow one to discriminate between them.

Then, we exploit the power of cosmic shear tomography to constrain the model parameters. We separate the Euclid distribution of sources into ten redshift bins, each containing roughly the same number of sources. In the present study, we neglect the contribution of intrinsic alignments of the source shapes. We find that Euclid is able to tightly constrain the parameters of these $f(R)$ models, as shown in Tables~1-2. Since the choice of the largest angular scale entering in the computation of the Fisher matrix is of primary importance, we perform the Fisher analysis for more $\ell_\mathrm{max}$'s. In particular, we use a conservative value of $1000$, an optimistic value of $5000$ and a bin-dependent setting, which increases the maximum angular wavenumber for distant shells and reduces it for nearby shells.

Finally, we compute the Bayesian expected evidence (e.g. \cite{Trotta:2005ar,Heavens:2007ka}) for the HS model over the $\Lambda$CDM model as a function of the extra parameter $\nhs$. This can be done because the \lcdm\ model is formally nested in the HS model, and the latter is equivalent to the former when $\nhs=0$. In this case, the two other extra parameters, $c_1$ and $c_2$, are linked together by the request of reproducing a $\Lambda$-like term in the gravitational Lagrangian. The expected evidence clearly shows that the Euclid survey data will unquestionably favour the HS model if any value $\nhs\gtrsim0.02$ is measured -- according to either the optimistic or the bin-depending analysis.

\acknowledgments We thank the referee for providing very insightful comments. SC and AD thank Bhuvnesh Jain for relevant suggestions on the use of cosmic shear tomography. SC acknowledges Research Grants funded jointly by Ministero dell'Istruzione, dell'Universit\`a e della Ricerca (MIUR), by Universit\`a degli Studi di Torino and by Istituto Nazionale di Fisica Nucleare within the \textit{Astroparticle Physics Project} (MIUR contract number: PRIN 2008NR3EBK). SC and AD gratefully acknowledge partial support from the INFN grant PD51. VFC is funded by the Italian Space Agency. This research has made use of NASA's Astrophysics Data System.

\bibliographystyle{JHEP}
\bibliography{/Users/montecristo/Documents/LaTeX/Bibliography}

\end{document}